\documentclass[3p,review,pdftex,numbers]{elsarticle}

\usepackage[T1]{fontenc}
\usepackage{hyperref}

\usepackage{appendix}
\usepackage[utf8]{inputenc}
\usepackage{array}
\usepackage{amssymb, amsmath, amsthm}
\usepackage{graphicx}
\usepackage{lmodern,url}
\usepackage{makecell} 
\usepackage{cancel}
\usepackage{multirow}
\usepackage{microtype}
\usepackage{lineno}
\usepackage{xspace}
\usepackage{xcolor}
\usepackage{soul}
\usepackage{siunitx}
\usepackage{todonotes}
\usepackage{booktabs}
\usepackage{csquotes}
\usepackage[ruled,vlined]{algorithm2e}

\usepackage{mathdots}
\usepackage{subcaption}
\newcommand{\eg}{\emph{e.g.}\xspace}

\newcommand{\dis}{\displaystyle}


\graphicspath{{Oficial Figures/}}

\definecolor{beaublue}{rgb}{0.74, 0.83, 0.9}
\definecolor{mygreen}{rgb}{0,0.5,0}
\newcommand{\Ri}[1]{{#1}} %

\newcommand{\Rii}[1]{{#1}} %

\begin{document}
%

\begin{frontmatter}
\title{On the heterogeneous spread of COVID-19 in Chile}

\author[DIQBM,CEBIB]{Danton Freire-Flores\corref{cor2}}
\author[DIQBM,CEBIB]{Nyna Llanovarced-Kawles\corref{cor2}}
\author[CEBIB,ICDB]{Anamaria Sanchez-Daza\corref{cor}}\ead{ana.sanchez@ing.uchile.cl}
\author[DIQBM,CEBIB]{\'Alvaro Olivera-Nappa}

\cortext[cor]{Corresponding Author}
\cortext[cor2]{These authors contributed equally to this work}
\address[DIQBM]{Department of Chemical Engineering, Biotechnology, and Materials, Universidad de Chile, Beauchef 851, 8370448 Santiago, Chile.}
\address[CEBIB]{Centre for Biotechnology and Bioengineering, Universidad de Chile, Beauchef 851, 8370448 Santiago, Chile.}
\address[ICDB]{Institute for Cell Dynamics and Biotechnology, Beauchef 851, 8370456, Santiago, Chile.}
\date{\today}

\begin{abstract}

\Ri{Non-pharmaceutical interventions (NPIs)} have played a crucial role in controlling \Ri{the spread of COVID-19}. \Ri{Nevertheless, NPI efficacy varies enormously between and within countries, mainly because of population and behavioral heterogeneity.}
In \Ri{this} work, we adapted a \Ri{multi-group SEIRA} model to \Ri{study} the spreading dynamics of COVID-19 in Chile, \Ri{representing} geographically separated regions \Ri{of the country by different groups. We use} national mobilization statistics to estimate the connectivity \Ri{between regions} and data \Ri{from governmental repositories to obtain \Rii{COVID-19} spreading and \Rii{death rates in each region}}. \Ri{We then} assessed the effectiveness of different \Ri{NPIs by studying the temporal evolution of the reproduction number $R_t$. Analyzing data-driven and model-based estimates of $R_t$, we found a strong coupling of different regions,} highlighting the necessity of organized \Rii{and} coordinated actions to control \Ri{the spread of SARS-CoV-2. Finally}, we evaluated different scenarios to forecast the evolution of COVID-19 in the most densely populated regions, \Ri{finding} that the early lifting of \Ri{restriction} probably \Ri{will} lead to novel outbreaks.

\end{abstract}

\begin{keyword}
COVID-19 \sep SARS-CoV-2 \sep Chile \sep Inverse problems \sep Reproduction number \sep Epidemiological model \sep SEIRD model
\end{keyword}
\end{frontmatter}

\section{Introduction}

The Severe Acute Respiratory Syndrome Coronavirus 2 (SARS-CoV-2) is the seventh reported coronavirus that can infect humans \citep{sixcovs,sevenCOV}. As a consequence of the fast global spread and severe effects of the infectious disease caused by SARS-CoV-2, COVID-19 was declared a pandemic by the World Health Organization on 11 March 2020, resulting in 214 countries affected to date \citep{Worldometerscountries,mesesref}. 
After \Ri{more than one year} of living in a pandemic world and despite scientific efforts, \Ri{effective}  \Ri{treatments for COVID-19 are not yet available. 
Even though \Ri{several countries are deploying vaccination plans worldwide}, uncertainties related to vaccine efficacy and uptake suggest the necessity of keeping non-pharmaceutical interventions (NPIs) in place for preventing excess deaths and the emergence of escape variants \citep{moore2021vaccination,bauer2021relaxing,contreras2021risking}.}

Since the pandemic began, \Ri{researchers have been proposing different epidemiological models to evaluate and forecast} the evolution of the disease. A significant part of those models derives from the renowned SIR model proposed \Ri{by Kermack and McKendrick in 1927} \citep{kermack1927contribution}, which compartmentalizes the population exposed to the virus in the variables susceptible ($S$), infected ($I$) and Recovered ($R$), whose interaction determines the evolution of the disease over time.  \Ri{Even though helpful, SIR models rely on several hypotheses that are rather hard to meet \citep{hethcote1985stability}, most of them related to homogeneity \citep{tolles2020modeling}. SIR models assume a "perfect mixing", where individuals are equally likely to meet, and from a continuous perspective, fractions of them do all the time. Individuals are assumed to have the same transition rates, translating to the same probability of being infected and the same average time to recover. Furthermore, SIR models do not include a latent period, one of the signature characteristics of COVID-19 \citep{he2020temporal}.}
Direct extensions of the SIR model include extra compartments to represent variables of epidemiological interest, as SEIR models, differentiating exposed individuals that are not yet infectious, SIRD, differentiating deaths, and SEIRA models, considering asymptomatic carriers, among others \citep{gondim2020seird,lalwani2020predicting,boujakjian2016modeling,cooper2020sir, weiss2013sir,postnikov2020estimation,SIRlimitations,contreras2020low}.

\Ri{Viral spread} depends not only on its biological \Ri{properties} but also on the behavior \Ri{and susceptibility of the population where it propagates} \citep{GeoImpact}. Therefore, a more realistic analysis of the propagation of the disease in a heterogeneous population usually requires models that include the interaction between the different sub-populations \citep{hethcote1985stability,kong2016,hethcote2009basic}. These sub-populations interact dynamically with each other, and the spreading modes are not necessarily isotropic \citep{contreras2020multigroup}. \Ri{Furthermore, our understanding of viral spread is greatly determined by data availability and quality, which has been proven to have significant delays \cite{contreras2020statistically}, weekly modulation \cite{dehning2020science}, and even weekend effects \cite{constantinesco2020spectral,bird2020now}.}

The case of Chile is an example of countries where typical SIR models would not work; its geopolitical centralization isolates and scatters the different regions that behave as independent populations with varying rules of interaction \citep{chileregs, contreras2020multigroup,contreras2020statistically}. Besides, the profound economic inequality as reported by the Gini's index of the world bank \citep{WorldbankGini} among different social classes constitutes both physical and behavioral anisotropy for the spread of COVID-19. \Ri{Chile has 16 regions with non-homogeneous connectivity, represented in the modeling by the interaction matrix ($\Phi$)}, in which each module can be weighted. This matrix is not necessarily symmetric, it considers the fractions of the communities that effectively interact, and it is modified by the spreading rate ($\beta$) of each sub-population according to the restriction measures implemented by the government to every community, representing the effective interaction in a given period.

In this work, we \Ri{modified} the multi-group SEIRA model presented \Ri{in} \cite{contreras2020multigroup} to \Ri{study} the COVID-19 spreading dynamics in Chile. \Ri{We incorporate a compartment for the deaths and design a tailored parameter fitting methodology to fit the parameters governing the dynamics. We incorporate real-world data for estimating connectivity matrices and solve the inverse problem for parameter fitting using data from official sources.} We assessed \Ri{the effectiveness of NPIs} by studying how parameters evolved in the different epidemiological periods. \Ri{We also study the differences between data-driven and model-based estimates of the reproduction number to evaluate whether the observed dynamics are driven by the local \Ri{behavior} or by the coupling with other regions}. 
By a Monte Carlo-inspired procedure \citep{montecarlo}, we estimated parameter variability and provided a statistically-based forecast of the pandemic's evolution in Chile.
We also \Ri{study} three \Ri{future} scenarios \Ri{assuming different levels} of social distancing, which allow us to predict the effect of the various health policies implemented by the government\Ri{; in consequence, authorities can take better decisions to prevent the spreading \citep{chimmula2020time}.}

\section{Methodology}

\subsection{Overview}

\begin{figure}[ht!]
    \centering
    \includegraphics[width=15cm]{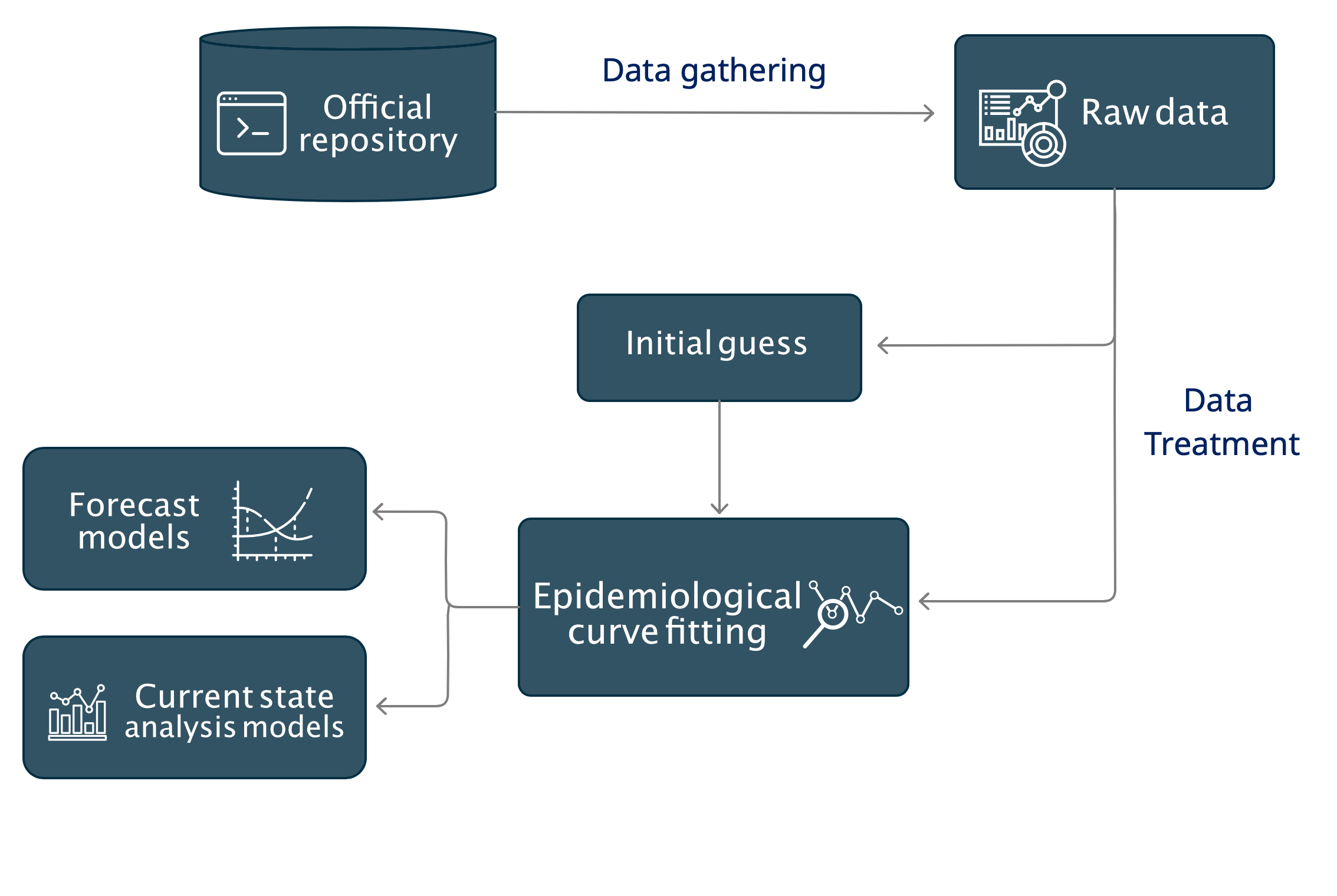}
    \captionsetup{justification=centering}
    \caption{Schematic representation of the inverse problem-solving pipeline.}
    \label{ProblemaInverso}
\end{figure}

\Ri{We detail the implementation workflow} of the proposed SEIRD multi-group model in Figure~\ref{ProblemaInverso}. \Ri{We first collect and pre-process official} data of the spread of COVID-19 in Chile \Ri{from} local authorities \citep{MINCIENCIA}. \Ri{The interaction and mobility data required for modeling the connectivity between the sixteen regions of Chile is estimated from national statistics \citep{censo2002,censo2012,censo2017} and projections for 2020 \citep{proyeccionespob}. We obtained default and initial parameter values for the model} from regional reports and literature. Then, \Ri{we study the temporal evolution of the spreading and \Rii{death} rate in each region, using a combination of} the \textit{simulated annealing} \Ri{and gradient descent algorithms} to minimize the error existing between the reported (raw) and simulated curves. Finally, using these values and based \Ri{ on different social distancing measures, various} possible scenarios are presented for active infected and deceased cases within an established temporal horizon.
  
\subsubsection{\Ri{Interaction dynamics} between different regions}

We summarize the interaction structure between the different sub-populations considered in the SEIRD model in a connection matrix ($\Phi$). In \Ri{this} matrix, entries $\left(\Phi\right)_{i,j}$ represent the fraction of individuals from the $i$th sub-population that moves to the $j$th. Note that $\Phi$ is not necessarily symmetric, as the \Ri{flux of individuals moving from $i$ to $j$ may be different from the ones from $j$ moving to $i$. An example of this situation is the migration to centralized regions, where a considerable fraction of individuals move to, but there is no considerable migration in the other way around}. Therefore, $\left(\Phi\right)_{i,j}\neq \left(\Phi\right)_{j,i}$.
In our model, $\Phi$ is also modulated by the local spreading rates $\beta_i$, thereby including governmental restrictions valid in the different periods considered.

For each region $i$, we estimated $\left(\Phi\right)_{i,j}$ as the fraction of individuals over fifteen \Ri{years old} from $i$ temporally moving to $j$ because of work or studies, \Ri{based on national mobility data \cite{censo2002,censo2012,censo2017,proyeccionespob}. The} displacement matrix $\Phi$ comprises factors $e_i$ and $C_{ij}$\Ri{;} The value $e_i$ represents the fraction of the population that leaves the region, defined as the number \Ri{of} people \Ri{from} region $i$ that travels to another region, divided by its total population $n_{\text{hab},i}$: 

\begin{equation}
    e_{i} = \frac{\dis\sum_{j\neq i} \text{individuals from Region}\,i}{n_i}
\end{equation}

\Ri{On the other hand, $C_{ij}$ represents the fraction of those individuals leaving $i$ that go to region $j$. Thus, we can estimate $\left(\Phi\right)_{ij}$ as:}

\begin{equation}
    \left(\Phi\right)_{ij} = e_i C_{ij},\qquad \left(\Phi\right)_{ii} = 1-\sum_{k\neq i} e_i C_{ik}.
\end{equation}


Given the geographical characteristics of Chile and its centralization, Regions do not have the same level of displacement between them (Table S2).

\subsection{Model Description}

\subsubsection{Equations}

\Ri{In our model, $i$ represents one of the sixteen regions of Chile, and we represent the spreading dynamics of SARS-CoV-2 using differential equations in a SEIRD compartmental model where the temporal scale is measured in days.}
Susceptible individuals $S_i$ can acquire the virus after an effective contagion from an infected individual and would be moved to the exposed compartment $E_i$. There, they would spend, on average, a time $t_c$ (latent period) until they become infectious $I_i$. After an average time $t_i$ (infectious period), infectious individuals are moved to a final recover $R_i$ or death $D_i$ compartment. The description, initial values, and sources for all the parameters in equations from (1) to (5) are shown in Table~\ref{tab:parametros}.

\begin{table}[hbt!]
   \captionsetup{justification=centering}
    \caption{Parameters description and initial values range assignment for the inverse-problem solution.}
    \label{tab:parametros}
    \centering
    \setlength{\arrayrulewidth}{0.5mm}
    \setlength{\tabcolsep}{3pt}
    \renewcommand{\arraystretch}{1.2}
\Ri{\begin{tabular}{clccc}
\toprule
 Parameter & Description & Value or range & Units & Source \\
\midrule
    $\alpha$ & Asymptomatic ratio of the population & 14 &  \SI{}{\%} & \citep{MINCIENCIA} \\
    $\beta_i$ & Spreading rate of the virus in \Ri{region} $i$ &  0.00 -- 0.30 &  \SI{}{days^{-1}} &\citep{contactrate} \\
        $\gamma$ & Recovery rate & 1/14 &  \SI{}{days^{-1}} &    \citep{gammaref} \\
    $\epsilon$ & Latent to infectious transition rate & 1/5 &  \SI{}{days^{-1}} & \citep{epsilonref,bar2020science,li2020substantial} \\
    $\theta_i$ & COVID-19 induced death rate in \Ri{region} $i$ & 0.00 -- 0.02 & \SI{}{days^{-1}} &        \citep{MINCIENCIA} \\
    $\xi$ & Factor of behavioral virulence of asymptomatics &        1 &   \SI{}{-} &       \citep{asintomaticos} \\
\bottomrule
\end{tabular}} 
\end{table}

Assuming that natural birth and death rate of the population can be neglected when compared with the COVID-19 induced death rates --which also occur in much shorter timescales--, \Ri{the} system of \Ri{differential} equations describing the dynamics is:

\begin{align}
    \frac{dS_i}{dt}  &=  -(1+(\xi-1)\alpha)\sum_{j=1}^n\beta_j\left(\frac{\Phi^{ij}S_i}{n_j^T}\right)\left(\sum_{k=1}^{n}\Phi^{kj}I_k\right) \label{dsfinal}\\
    \frac{dE_i}{dt}  &= (1+(\xi-1)\alpha)\sum_{j=1}^n\beta_j\left(\frac{\Phi^{ij}S_i}{n_j^T}\right)\left(\sum_{k=1}^{n}\Phi^{kj}I_k\right)\qquad-\epsilon_i E_i\\
    \frac{dI_i}{dt}  &= \epsilon_i E_i -(\gamma_i+\theta_i)I_i\label{dI}\\
    \frac{dR_i}{dt}  &= \gamma_i I_i\\
    \frac{dD_i}{dt}  &= \theta_i I_i
\end{align}

\subsection{Data treatment}

Raw data presented corresponds to that reported daily by the Government of Chile, available in \citet{MINCIENCIA}. 
This work analyzed official data reported daily by the Government of Chile \citep{MINSAL}, consisting in a total of 244 days until November \Ri{the 2nd, 2020}.

Regional-level daily data for each region $i$ starts in March 3rd \Ri{2020} and include new daily infected cases, total infected cases, and total deaths ($D_{i}$). However, due to a reporting inconsistency generated when the official reporting guidelines changed (June 1st), datasets contained systematic errors. We corrected those errors as explained in Supplementary Section~3. 
Using these corrected datasets (available as supplementary dataset), we calculate total recovered ($R_{i}$) and active infected cases ($I_{i}$), defining also variables accounting for the experimental time ($T_{\rm exp}$) and number of inhabitants per region ($n_{\rm hab}$).

For parameter fitting, we defined fifteen epidemiological periods for the pandemic progression in Chile, which limits matched relevant governmental measures detailed in Table~\ref{tab:periods} \citep{Lineadetiempo,estadoexcepcion,infoepide}.


\begin{table}[h!]
\small
\captionsetup{justification=centering}
    \caption{Epidemiological periods.}
    
\label{tab:periods}
\centering
\setlength{\arrayrulewidth}{0.5mm}
\setlength{\tabcolsep}{5pt}
\renewcommand{\arraystretch}{0.9}
\begin{tabular}{lllll}
\toprule
\textbf{Period} & \multicolumn{3}{c}{\textbf{2020 dates}} & \textbf{Remarkable government measures}                       \\ \midrule
I           & Mar 3 & -- & Apr 6      & \makecell[l]{3/3 First case reported,  3/15 Mandating Schools closure\\ 3/16 Borders were closed, 3/18 \textquote{State of exception$^{*}$} enacted \\ 3/21 First deceased person, 3/25 Lockdown in some Metropolitan Region areas}    \\ \midrule
II              & Apr 7& -- & Apr 21     & 4/19 Government call for a \textquote{new normality}          \\ \midrule
III             & Apr 22& -- & May 6     & 4/23 School were indefinitely closed                \\ \midrule
IV              & May 7& -- & May 21     & 5/13 Announcement of lockdown for 90\% of Metropolitan Region   \\ \midrule
V               & May 22& -- & Jun 5     & \makecell[l]{Health Ministry defines new COVID-19 case-reporting criteria:\\     - Expansion of the dead case criterion\\     - Active contagious case since the onset of symptoms\\     - Include positive PCRs in laboratory reports as active cases}   \\ \midrule
VI              & Jun 6& -- & Jun 20     & \makecell[l]{Change of Health Minister \\ 6/16 The \textquote{state of exception}$^{*}$ is extended for another 90 days\\ 6/17 31,412 previously dismissed cases were integrated into the official count\\ 6/18 Previously unnotified cases were added to the daily infections \\ list due to an inclusion criterion change } \\ \midrule
VII             & Jun 21& -- & Jul 05    & \makecell[l]{Chile reached sixth place in the total number of confirmed cases worldwide\\ 7/4 Some communes of the Metropolitan Region completed 100 days in quarantine}                                                                                          \\ \midrule
VIII            & Jul 6& -- & Jul 20     & 7/19 Announcement and enactment of the \textquote{Step by Step} lockdown release plan  \\ \midrule
IX              & Jul 21& -- & Aug 4     & 7/28 Lockdown release for districts in the Metropolitan and Valparaíso Regions   \\ \midrule
X               & Aug 6& -- & Aug 19     & 7/28 Transition step area broadened in the Metropolitan Region     \\ \midrule
XI              & Aug 20& -- & Sep 3     & 8/28 Several districts of Bío-Bío and Maule Regions went back to lockdown  \\ \midrule
XII             & Sep 4& -- & Sep 18     & Communes of O'Higgins, Magallanes Regions, among others, move back to lockdown             \\ \midrule
XIII            & Sep 19& -- & Oct 3     & \makecell[l]{The state of exception due to catastrophe will remain\\          in force by a presidential mandate for the next 79 days}    \\ \midrule
XIV             & Oct 4& -- & Oct 18     & First schools reopening                                                                                                                                                                                                                                                    \\ \midrule
XV              & Oct 19& -- & Nov 2     & 10/25 Chile's 2020 National Plebiscite                                                             \\ \bottomrule
\multicolumn{5}{l}{$^{*}$\footnotesize{In which the \Ri{government} may transcend the rule of law in the name of the public good.}}

\end{tabular}
\end{table}                                                      

\subsection{Initial guess and nested problems}

The set of initial values for the parameters (Supplementary Table~S3) are expected to reflect the behavior of the different population variables (SEIRD) in the different epidemiological periods, according to the confinement measures established by the government in every region. We perform a separate parameter fitting for every epidemiological period and could be extended to a higher number of periods, if necessary.

After obtaining the simulated curves for the SEIRD variables, using the values in Table~S3, we define which parameters will be estimated by the simulated annealing algorithm, before proceeding to the formal parameter fitting. We considered the percentage of asymptomatic patients to be constant across regions, based on governmental reports \citep{MINSAL}. Recovery and exposure rates are considered uniform at the country level as the Chilean health system have not been overwhelmed to date. Therefore, we obtain in the parameter fitting the local spreading and death rates ($\beta_i$ and $\theta_i$ respectively). We also provide freedom to the initial conditions \Ri{of the first epidemiological period}, leaving them to be \Ri{determined} in the parameter fitting procedure.

\Ri{We numerically obtain the solution of the SEIRD model (equations (1)--(5)) for every parameter combination evaluated throughout the fitting process}. Subsequently, we quantify a measure proportional to the mean squared error (MSE) between simulated curves and raw data reported for each region, combined in the cost functional $J^k$ being minimized \Ri{in the $k$'th epidemiological period}. The optimization algorithm selected was a combination of \textit{simulated annealing} and gradient descent.

Finally, the initial guess for initial condition of the SEIRD variables in each region was defined as follows: the initial value for the number of susceptible ($S_i(0)$) people is considered as the total population of the region, while $E_i(0) = I_i(0) = R_i(0) = D_i(0) = 0$, except in Maule Region, where $I_{\rm Maule}(0) = 1$ and $S_{\rm Maule}(0)=n_{\rm hab}-1$ because \Ri{this region  was where the first COVID-19 case in Chile was reported}.

\subsection{Parameter fitting strategy: Resolution of inverse problem}

We determined the set of \Ri{parameters} that best describes the \Ri{observed national dynamics (with regional resolution)} by minimizing \Ri{a} cost \Ri{functional} $J^k$ for each of the $k$ epidemiological periods $T_k = [t_{i}^{k},t_f^{k}]$ as described in Table~\ref{tab:periods}. This functional \Ri{accounts for the total MSE (mean squared error) between} the number of infected $I$, deaths $D$, and recovered $R$ cases and raw data \Ri{obtained from official repositories~\cite{MINSAL}}. \Ri{Structurally, $J^k$ includes the three contributions:}

\Ri{\begin{equation}
    J^k = J_I^k+J_R^k+J_D^k.
\end{equation}}

Because of \Ri{differences} in the number of inhabitants across regions, \Ri{without explicitly correcting it,} the error contributed by a small region will be less than that contributed by a large region. To avoid unrealistic solutions where small regions would be left aside because of the algorithm's blind drive to minimize the error, we included a weighting factor. Each contribution of each region was weighted by $w_X^{i,k}$, \Ri{a} factor directly proportional to the ratio between \Ri{the total infected, recovered or deceased calculated cases for the all country in the last day of the epidemiological period}, and the \Ri{respective} number of infected, recovered or deceased \Ri{calculated cases in the $i$ region the same day}. 

\Ri{\begin{align}
    J_X^k  &= \sum_{i=1}^{16}\sum_{t\in[t_{i}^{k},t_f^{k}]}w_X^{i,k}\left(X^{\rm mod}_{i}\left(t\right)-X^{\rm data}_{i}\left(t\right)\right)^2 \\
    w_X^{i,k} &= w_X\left(\frac{\sum_{i=1}^{16}X^{\rm mod}_{i}\left(t_f^{k}\right)}{X^{\rm mod}_{i}\left(t=t_f^{k}\right)}\right)^{1.7},
\end{align}}

\Ri{where the 1.7 exponent in $w_X^{i,k}$ was empirically set for balancing the contributions of large and small regions, and the correction} factor $w_X$ connects the different contributions in functional $J_X$. Given the difference in magnitude of deceased in comparison with active infected and recovered cases we define $w_I = w_R = 1$ and $w_D = 10^2$.

\Ri{We then find the optimal parameters for the fit as the argument of a minimization problem, numerically \Rii{minimizing} value of $J_k$ for each epidemiological period. Noteworthy, each epidemiological period described in Table~\ref{tab:periods} has at least 14 data points ranging in times $[t_{i}^{k},t_f^{k}]$ corresponding to daily levels of infection, recoveries, and deaths, thus constituting for each region (and per period) a total of 42 independent measures. On the other hand, we only have to fit two parameters per region and per epidemiological period (namely, $\beta_i^k$ and $\theta_i^k$), thus not risking overfitting.}

\Ri{The implementation of the mathematical model and resolution of the parameter-fitting inverse problem was performed in Matlab version R2018a.}

\subsection{Variability assessment and forecasting}

To obtain reliable values for the selected parameters, Monte Carlo simulations \citep{montecarlo} were performed according to the methodology described in \citet{contreras2020novel} (\Rii{$N=450$}). \Ri{In these simulations, we induce a $\pm 5\%$ white noise in the input data, and fit the parameters using this mildly noisy signal. Doing so, we aim to minimize the contribution of potential errors underlying the data. In that way, this experiment is also a sensitivity analysis of the method to the data.} The inverse problem for curve-fitting is solved individually, resulting in distributions of parameters. We statistically obtained average population parameters and their variability from those distributions. 

Once obtaining values for both parameters and their uncertainty, we evaluate different forecast scenarios for the evolution of COVID-19 pandemic Chile.

\section{Results \Ri{and discussion}}

\subsection{Monte Carlo simulations}

As a result of Monte Carlo multiple-simulations experiment performed we obtained distributions for both $\beta_i$ and $\theta_i$ parameters in each region, and subsequently used them to numerically solve our model. Here we focus on  active infected cases ($I$) and total deaths ($D$), since these we believe \Ri{are of} greater importance for \Ri{policymakers}.

Tables \ref{tab:betas} and \ref{tab:thetas} show the median values obtained for $\beta$ and $\theta$, respectively, with a 95\% confidence interval in the Metropolitan, Valparaíso and Bío-Bío Regions \Ri{in each epidemiological period}. Noteworthy, these regions \Ri{concentrate} the largest number of inhabitants in the northern, central, and southern zones of the country\Ri{, adding up to} almost 60\% of the total population of Chile. We observe that the highest values of $\beta$ were in the first epidemiological period, where no social distancing measures had been established, followed by the IV epidemiological period, for Metropolitan, Valparaíso, and Bío-Bío Regions \Ri{--- which was approximately} three weeks \Ri{after} the government announced \textquote{safe return to work} or \textquote{\textit{the new normality}} in the Metropolitan Region ---. \Ri{We observe a change point in the general population behavior after the announcement, which was delayed to the IV epidemiological period because of the disease timeline (latency, incubation, recovery time) and significant delays in testing \cite{contreras2020statistically}}.
A higher rate of infections subsequently spread from Metropolitan Region to the regions with the highest rate of transfers: Valparaíso and Bío-Bío. In each epidemiological period, $\beta_i$ values do not present a high variability, which is reflected in the \Ri{narrow} 95\% confidence intervals. The periods in which there is a higher variability correspond to those with higher $\beta$ values, so the number of infections was triggered and the fitting becomes more challenging.

\begin{table}[ht!]
   \captionsetup{justification=centering}
    \caption{\Rii{$\beta$ [\SI{}{days^{-1}}] median values with 95\% confidence intervals for each epidemiological period for Metropolitan, Valparaíso, and Bío-Bío Regions.}}
    \label{tab:betas}
    \centering
    \setlength{\arrayrulewidth}{0.5mm}
    \setlength{\tabcolsep}{8pt}
    \renewcommand{\arraystretch}{1.2}
\begin{tabular}{ccc|cc|cc}
\toprule

       Epid. Period & \multicolumn{2}{c}{Valparaíso} & \multicolumn{2}{c}{Metropolitan} & \multicolumn{2}{c}{Bío-Bío} \\
       \hline
          & Median & CI    & Median & CI    & Median & CI \\
    I     & 0.98  &     [0.95--0.98]    & 1.03  &  [1.01--1.04] & 1.10  & [1.06  1.11] \\
    II    & 0.05  &     [0.05--0.05] & 0.14  &     [0.14--0.15] & 0.01  &     [0.01--0.01] \\
    III   & 0.11  &     [0.11--0.12] & 0.29  &     [0.29--0.29] & 0.01  &     [0.01--0.01] \\
    IV    & 0.21  &     [0.21--0.22] & 0.19  &     [0.19--0.20] & 0.26  &     [0.25--0.28] \\
    V     & 0.10  &     [0.10--0.10] & 0.14  &     [0.14--0.15] & 0.14  &     [0.14--0.15] \\
    VI    & 0.09  &     [0.09--0.09] & 0.06  &     [0.06--0.06] & 0.08  &     [0.07--0.08] \\
    VII   & 0.05  &     [0.05--0.05] & 0.05  &     [0.05--0.05] & 0.14  &     [0.14--0.15] \\
    VIII  & 0.05  &     [0.05--0.06] & 0.01  &     [0.01--0.01] & 0.04  &     [0.04--0.04] \\
    IX    & 0.05  &     [0.05--0.05] & 0.02  &     [0.02--0.02] & 0.06  &     [0.06--0.06] \\
    X     & 0.07  &     [0.07--0.07] & 0.12  &     [0.12--0.13] & 0.10  &     [0.10--0.11] \\
    XI    & 0.09  &     [0.09--0.09] & 0.01  &     [0.01--0.01] & 0.10  &     [0.10--0.12] \\
    XII   & 0.06  &     [0.06--0.07] & 0.08  &     [0.08--0.09] & 0.05  &     [0.05--0.05] \\
    XIII  & 0.03  &     [0.03--0.03] & 0.03  &     [0.03--0.04] & 0.07  &     [0.07--0.07] \\
    XIV   & 0.06  &     [0.06--0.06] & 0.08  &     [0.08--0.08] & 0.07  &     [0.07--0.07] \\
    XV    & 0.05  &     [0.05--0.05] & 0.07  &     [0.07--0.07] & 0.10  &     [0.10--0.10] \\
    
\bottomrule
\end{tabular}     
\end{table}

\begin{table}[h!]
   \captionsetup{justification=centering}
    \caption{\Rii{$\theta$ [\SI{}{days^{-1}}] median values with 95\% confidence intervals for each epidemiological period for Metropolitan, Valparaíso and Bío-Bío Region.}}
    \label{tab:thetas}
    \centering
    \setlength{\arrayrulewidth}{0.5mm}
    \setlength{\tabcolsep}{8pt}
    \renewcommand{\arraystretch}{1.2}
    \begin{tabular}{ccc|cc|cc}
\toprule
   Epid. Period & \multicolumn{2}{c}{Valparaíso} & \multicolumn{2}{c}{Metropolitan} & \multicolumn{2}{c}{Bío-Bío} \\
\hline
 
          & Median & CI    & Median & CI    & Median & CI \\
    I     & 0.0029 &     [0.0027--0.0033] & 0.0031 &     [0.0027--0.0033] & 0.0031 &     [0.0027--0.0033] \\
    II    & 0.0024 &     [0.0023--0.0027] & 0.0024 &     [0.0023--0.0027] & 0.0010 &     [0.0009--0.0012] \\
    III   & 0.0039 &     [0.0036--0.0044] & 0.0033 &     [0.0028--0.0033] & 0.0020 &     [0.0018--0.0022] \\
    IV    & 0.0028 &     [0.0027--0.0033] & 0.0027 &     [0.0027--0.0027] & 0.0012 &     [0.0009--0.0014] \\
    V     & 0.0019 &     [0.0018--0.0022] & 0.0018 &     [0.0018--0.0018] & 0.0020 &     [0.0018--0.0022] \\
    VI    & 0.0018 &     [0.0018--0.0022] & 0.0018 &     [0.0018--0.0018] & 0.0008 &     [0.0005--0.0008] \\
    VII   & 0.0019 &     [0.0018--0.0022] & 0.0027 &     [0.0027--0.0027] & 0.0010 &     [0.0009--0.0014] \\
    VIII  & 0.0018 &     [0.0018--0.0022] & 0.0027 &     [0.0027--0.0027] & 0.0009 &     [0.0009--0.0009] \\
    IX    & 0.0028 &     [0.0027--0.0033] & 0.0027 &     [0.0027--0.0031] & 0.0009 &     [0.0009--0.0012] \\
    X     & 0.0020 &     [0.0018--0.0022] & 0.0018 &     [0.0018--0.0022] & 0.0009 &     [0.0009--0.0013] \\
    XI    & 0.0030 &     [0.0027--0.0033] & 0.0018 &     [0.0018--0.0022] & 0.0009 &     [0.0009--0.0014] \\
    XII   & 0.0020 &     [0.0018--0.0022] & 0.0018 &     [0.0018--0.0022] & 0.0009 &     [0.0009--0.0014] \\
    XIII  & 0.0020 &     [0.0018--0.0022] & 0.0019 &     [0.0018--0.0022] & 0.0019 &     [0.0018--0.0022] \\
    XIV   & 0.0031 &     [0.0027--0.0033] & 0.0020 &     [0.0018--0.0021] & 0.0012 &     [0.0009--0.0014] \\
    XV    & 0.0032 &     [0.0027--0.0033] & 0.0019 &     [0.0018--0.0022] & 0.0021 &     [0.0018--0.0022] \\

\bottomrule
\end{tabular}     
\end{table}

These analyses are extended to each of the 13 remaining regions of the country, reporting median values of $\beta_i$ and $\theta_i$ ($ i = 1 ... 16 $) for each region and epidemiological period (Supplementary Tables S4 and S5). \Rii{Raw data (parameter values for each realization of the Monte Carlo simulation) for all regions and epidemiological periods are reported as separate Supplementary Files for $\beta$ and $\theta$. The death rate $\theta_i$ should not be mistaken by the Infection Fatality Ratio (IFR), as the latter represents the fraction of individuals who die after being infected. If we focus on the infectious compartment $I_i$,  the transitions rates to recovery or death are given by $\gamma$ and $\theta_i$, being $\gamma=1/14$ at least one order of magnitude larger than $\theta_i$ (cf. Table~\ref{tab:thetas}). Estimating the IFR as $\frac{\theta_i}{\gamma+\theta_i}$, and using the inferred values of theta for all epidemiological periods and regions, we obtain a median IFR of \SI{1.56}{\%}, which agrees well with official data  \citep{reporteminsal}. A more detailed analysis is provided in the Supplementary Materials, Section S4}.

The NPI (non-pharmaceutical interventions) agenda of the government was different for each region, and the main criterion for enacting them was the reported new cases. We observe that the spreading rate $\beta$ remained high in some regions during the first epidemiological periods before decreasing, due to the time required for \Ri{evidencing} the effects of lockdowns and delays associated with the disease progression.
In those regions reporting the first COVID-19 cases in Chile, the spreading rate $\beta$ decreased faster because of the earlier establishment of NPIs. An example is the O'Higgins Region, which has the lowest average value for the fifteen epidemiological periods of spreading rate. Regarding the \Rii{death rate} $\theta$, it \Ri{remained} relatively low in all regions since the hospital capacity at the country level for the correct care of critical patients was not exceeded, and \Ri{also due} to the efficiency in the inter-regional transfer of patients.

\subsection{\Ri{COVID-19 regional spreading dynamics}} 

\begin{figure}[!h]
\centering
\captionsetup{justification=centering}
\includegraphics[scale=.28]{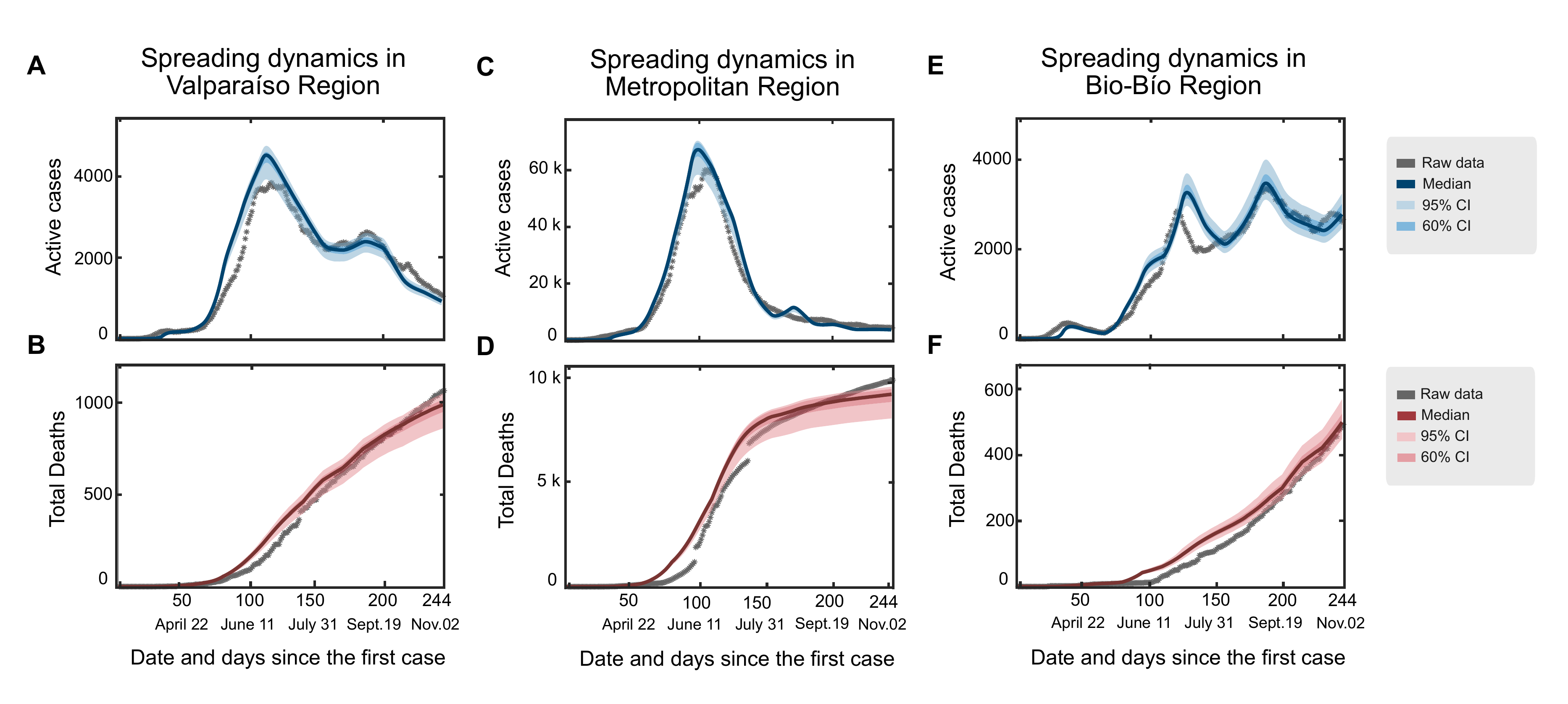}
\caption{Simulated curves of active infected (top) and deceased (bottom) cases for the Valparaíso (A,B), Metropolitan (C,D) and Bío-Bío (E,F) Regions, along with 60\% and 95\% confidence intervals. Raw data for the evaluated time-frame is also presented (grey dots).}
    \label{fig:ID_Regs}
\end{figure}

Using the different sets of parameters obtained in the Monte Carlo experiment, the simulated curves for the active infected cases (I) and deaths cases (D) are projected in each region over time with a two-level calculated confidence interval (60\% and 95\%). Figure~\ref{fig:ID_Regs} shows both the simulations and raw data for the Metropolitan, Valparaíso, and Bío-Bío Regions.  

We observe that simulated curves are in good agreement with raw data, which always remain within the confidence intervals and present consistent trends. The color intensity of these zones varies depending on the confidence intervals percentage (95\% and 60\%, light and dark shadows, respectively), calculated from the $\beta$ and $\theta$ distributions obtained in the Monte Carlo experiment. 

\Ri{We observe a peak in the active cases in the Metropolitan Region} between June and July, which corresponds to an increase in the number of deaths cases on the same \Ri{period}. \Ri{Raw} data for the active infected cases in the Valparaíso and Bío-Bío Regions has two peaks. The first one corresponds to the date where active cases in the Metropolitan Region were at their peak, so due to the high number of transfers between these regions, these active cases act as vectors of infection in the destination regions, thus increasing infected cases and therefore deceased. In contrast, the second one is related to the fact that the measures adopted by the government are sectorized for each Region. For Valparaíso and Bío-Bío Regions, the isolation measures were relaxed before than for the Metropolitan Region, resulting in a second increase rate of the active infected cases. \Ri{Curves for the other regions are presented in the supplementary material (Figure 1 to Figure 13, Supplementary). 
As the time-frame set for each epidemiological period does not necessarily match the temporality of the different measures enacted in each region, some parts of the raw data drift from the median in certain epidemiological periods. Thus, it is crucial to consider the actual spreading dynamics between geographical Regions to represent more appropriately the scenario in each Region.} This could \Ri{also} be solved by further reducing the number of days per period but risking the possibility of overfitting to raw data.

\begin{figure}[!h]
\centering
\captionsetup{justification=centering}
\includegraphics[width=15cm]{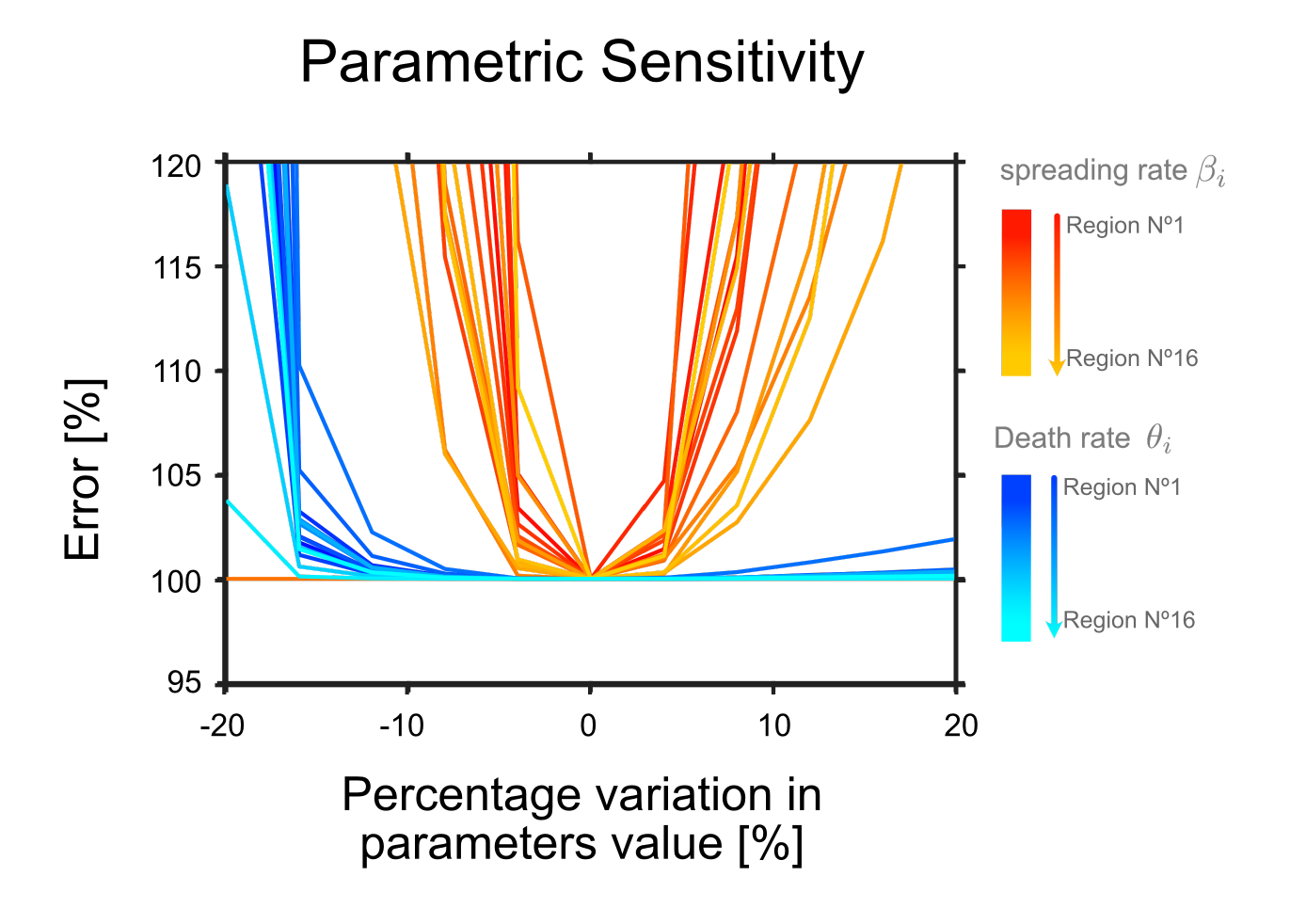}
\caption{\Ri{Sensitivity analysis for the last epidemiological period. The red scale represents the variations in $\beta$ values, from the first region (red) to the last (yellow), while the blue scale represents the variations in $\theta$ values, from the first region (blue) to the last one (cyan). }}
    \label{fig:AS}
\end{figure}

The proposed SEIRD model was able to adjust well to the data in both low-population and heavily populated regions, showing the relevance \Ri{of the 1.7} factor described in the \textit{parameter fitting strategy} section. Higher values for this adjustment factor result in a better adjustment in the small regions, but at the cost of a less rigorous adjustment in the larger regions. In contrast, with values lower than 1.7, the opposite occurs. \Ri{In Figure~\ref{fig:AS}, we present the results of a sensitivity analysis of the fitted set to perturbations of $\left(\pm 20\%\right)$ in the parameters obtained in for last epidemiological period. } 

\Ri{We observe that, when making these variations in each of the 32 parameters obtained (16 $\beta$ and 16 $\theta$) the calculated error increases, thus confirming that the parameters obtained by solving the inverse problem effectively minimize the error existing between the simulated curves and the experimental ones.}

\subsection{On the values of $R_t$ in coordinated government measures}

The Effective Reproduction Number $R_t$ represents the number of persons a single infected individual might infect, in a population that is aware of the disease representing the viral spread rate of the virus, and varies depending on the policies implemented by the government, such as quarantines \citep{perasso2018introduction}. 
Based on the parameters obtained from the simulations, it is possible to calculate the value of $R_t$ as the ratio between the spreading rate $\beta$ and the recovery plus deaths rate $(\gamma + \theta)$, as demonstrated in \citet{Rt_calculation}. This approach allows us to obtain the effective reproduction number driven only by the local population's behavior, decoupled from other regions:

\begin{equation}
    R_t = \frac{\beta_i}{\theta_i+\gamma_i}.
\end{equation}

We obtain a data-driven value for the observed $R_t$, $R_t^{\rm obs}$, adapting the methodology presented in \citet{contreras2020R0,medina2020country}:

\begin{equation}
    R_t^{\rm obs} = \frac{\Delta I}{\Delta R + \Delta D}+1
\end{equation}

In particular, both values do not necessarily need a match, as they represent -slightly- different quantities. The observed $R_t^{\rm obs}$ represents whether the overall trend is the spread or containment of the disease, purely driven by data, and is affected by testing and tracing governmental plans \citep{contreras2021challenges}. More extensive testing will uncover unnoticed infection chains, and also increase the numbers as the uncovering would be faster than the spread of the disease. On the other hand, the effective reproduction number $R_t$ accounts for local trends on contagion, disregarding whether those cases would be noticed or remain uncovered throughout the disease timeline.

\begin{figure}[h!]
    \centering
    \captionsetup{justification=centering}
    \includegraphics[scale=.34]{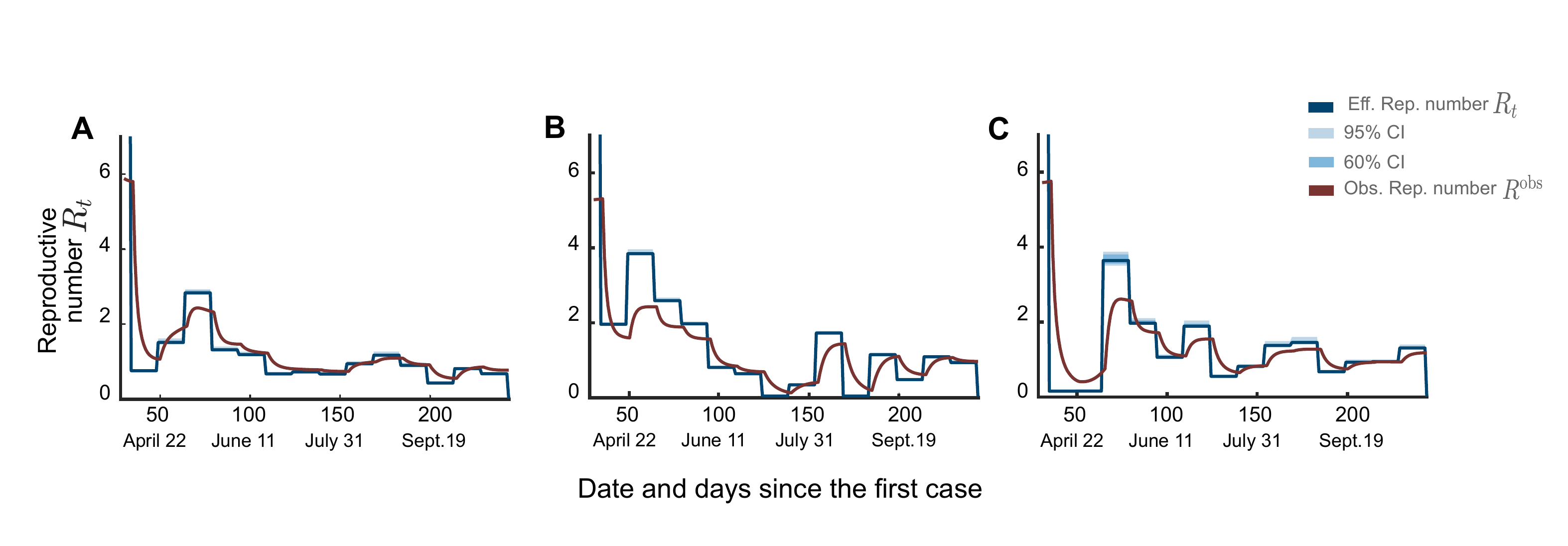}
    \caption{$R_{t}$ values calculated from raw data (red) and simulated data (blue) for the Valparaíso (A), Metropolitan (B) and Bío-Bío (C) Regions with \Rii{60\%} and \Rii{95\%} confidence interval.}
    \label{fig:Rts_Regs}
\end{figure}

In Figure \ref{fig:Rts_Regs} the $R_t$ values obtained from official sources and simulation results in the Metropolitan, Valparaíso, and Bío-Bío Regions are presented, in conjunction with an \Rii{95}\% confidence interval. As is shown, both $ R_t $ values reflect a similar behavior. However, they are not identical,  probably because there is a difference between the simulated and raw data due to the contribution of the inter-region movements in the infection rate, which agrees with the results obtained in the inverse problem-solving.  

These results for $R_t$ show the correlation between the imposed social distancing measures and the values of the Effective Reproduction Number, in the different regions. This value is higher in the III epidemiological period for the three presented regions and subsequently decreases as more restrictive measures are declared. The case of the Magallanes region (Figure 26, Supplementary Material) is of particular interest, in the absence of proper coordination around social distancing measures, a high $R_t$ number is observed in mid-August and consequently a second outbreak of infected assets, more significant than the first one. 

Projecting these $R_t$ values into the future, a decrease in the virus spreading rate and a reduction of the number of active infected would be expected, since a $R_t$ < 1 implies a slow rate of spread and the outbreak size would decay exponentially.

\subsection{Model Forecasting}

\begin{figure}[ht!]
    \centering
    \captionsetup{justification=centering}
    \includegraphics[scale=.5]{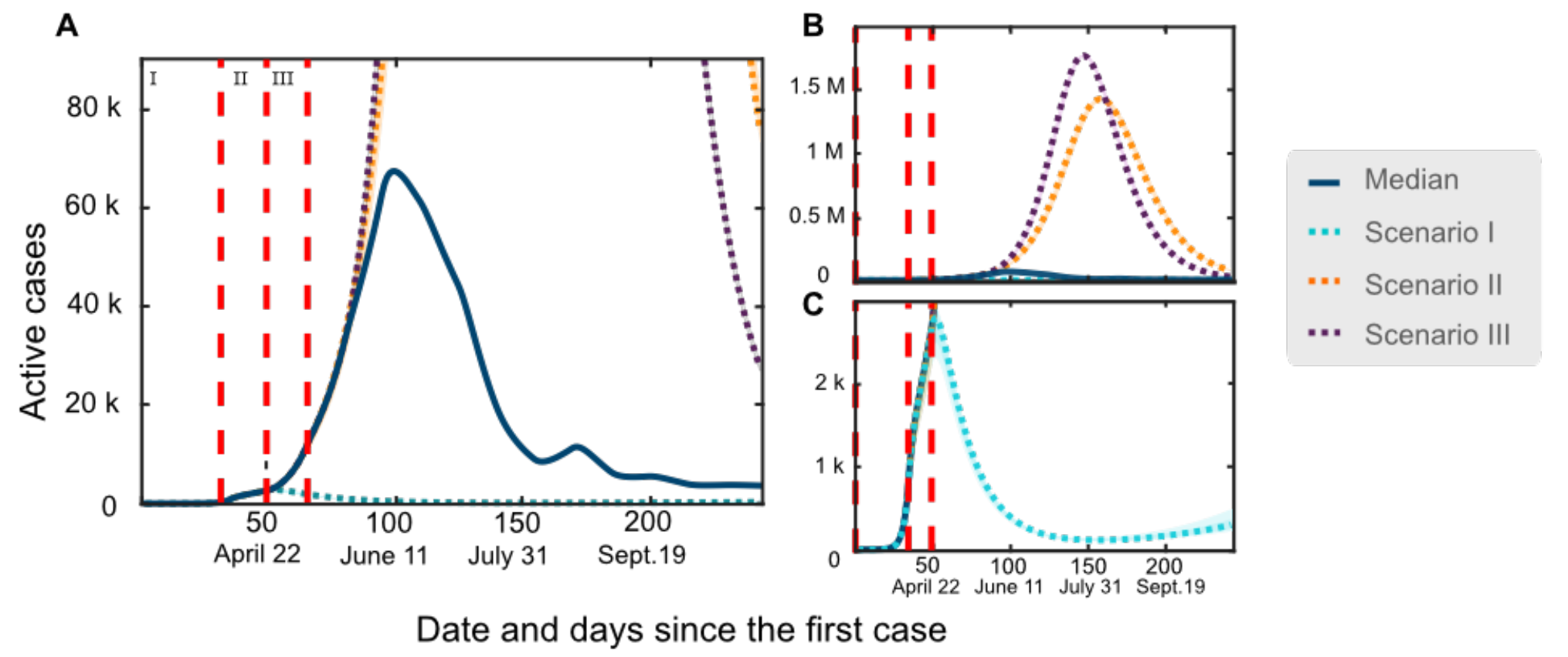}
    \caption{\textbf{Different simulated scenarios for active infected cases in the Metropolitan Region}. \Rii{\textbf{A:} Scenario I (cyan curve) corresponds to establishing a total quarantine from the first epidemiological period. In contrast, scenarios II (orange curve) and III (purple curve) represent the cases where no greater measures of social distancing took place as of periods II and III, respectively. The vertical dashed red lines indicate the date of periods I, II and III. Figures B and C correspond to different scales from figure A to visualize the scope of the curves.}}
    \label{fig:efMedidas}
\end{figure}

To evaluate the impact of governmental interventions, we simulated different scenarios aiming to project the contagion trends observed if those interventions did not take place. In Figure \ref{fig:efMedidas}, we present projections of infected cases in the Metropolitan Region, if no more restrictions were imposed after the II and III epidemiological periods.

Due to the importance of establishing containment measures as soon as possible to avoid an excessive spread of the virus, the first three epidemiological periods were chosen to evaluate immediate quarantine (I) and non-imposition of measures (II and III) scenarios. 

It is observed that in the scenario I the number of active infected cases reaches a peak of approximately 1500 (Figure \ref{fig:efMedidas}, C) cases, 30 days after the first reported case. The spreading rate ($\beta$) (Table \ref{tab:betas}) associated with this scenario corresponds to \Rii{0.05}, registered in the VII epidemiological period in which the metropolitan region was in total quarantine. On the other hand, in scenarios II and III active infections peak over a million cases between days 150 - 200 (Figure \ref{fig:efMedidas}, B). Scenario III \Ri{is} the one that presents a higher number of infections due mainly to two factors: the number of active infected cases at the beginning of the period and the spreading rate. In both factors, the values are higher in the third period than in the second. In addition to the spreading rate ($\beta$) (Table \ref{tab:betas}) in the third period being \Rii{0.29}, while in the second period it is \Rii{0.14}, a higher initial number of actively infected would facilitate the spread of the virus in the presence of a high spreading rate, explaining the behavior of the scenario II and III curves. 

Using the same strategy, \Ri{we analyzed} different future scenarios based on the \Ri{projected restrictions for} each region. On July 19th, the Government of Chile announced the beginning of the "Step by Step" plan to \Ri{lift the current social} distancing measures gradually. This plan involved the gradual opening in the different \Ri{districts and regions} of the country based on the contagion rate present in each one, the percentage of occupancy of ICU (intensive care unit) beds in hospitals, and the rate of PCR test positivity, among others. The steps are:

\begin{enumerate}
\item Quarantine: People cannot leave their homes.
\item Transition: People is allowed to leave the house with restrictions, only on weekdays.
\item Preparation: Individuals are free to move, but group gatherings are not permitted.
\item Initial Opening: Group gatherings are permitted, with a restricted number of people.
\item Advanced Opening: free group gatherings are permitted.
\end{enumerate}

Each region that enters a new stage of the plan presents greater freedom of movement and\Ri{, therefore, the number and intensity} of contacts between people \Ri{(and thereby the spreading rate} $\beta$) could increase. Consequently, a \textit{weighting factor} is assigned to each stage of the plan. This factor multiplies the $\beta$ value of the corresponding region to projecting the effects that the "Step by Step" program will have on the extrapolation in the SEIRD variables (Table S6) \citep{situacionchile}.

\begin{figure}[!h]
    \centering
    \captionsetup{justification=centering}
    \includegraphics[scale=.32]{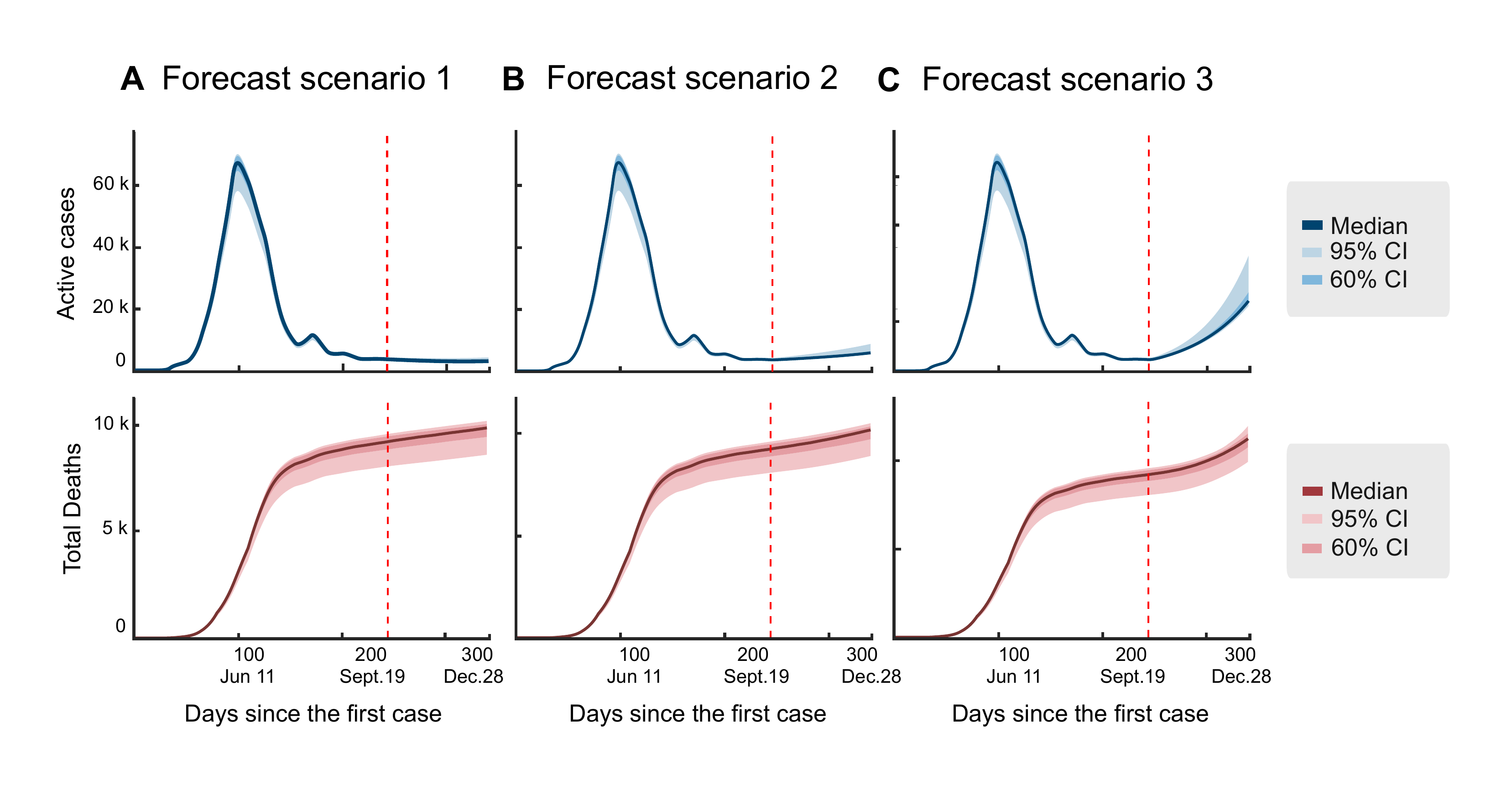}
    \caption{Different scenarios proposed for the Metropolitan Region. The first scenario (A) represents the evolution of the active infected (top) and death (bottom) cases maintaining a lockdown from November 2nd (vertical red line) until the end of 2020, while the second (B) and third scenario (C) represent the current situation and a massive reopening respectively.}
    \label{fig:ForecastRM}
\end{figure}

Figure \ref{fig:ForecastRM} present three different scenarios projected for the Metropolitan Region. These scenarios are projected in a time window until the end of 2020:

\begin{itemize}
  \item Scenario 1: The Metropolitan Region remain in stage 1 of quarantine. 
  \item Scenario 2: Current scenario. The Metropolitan Region advances to 4th phase  of preparation.
  \item Scenario 3: Limit scenario, in which the spreading rate $\beta$ increases to such an extent that large outbreaks and the second wave of massive infections occur.
  \end{itemize}

Based on the projections, we would expect a sustained decrease in case numbers if the quarantined regime continues. However, in the current scenario (initial opening phase), a second peak in contagion is expected. Nevertheless, this wave would be lower in magnitude compared with the one observed in the first half of 2020, and therefore the health system would also be able to handle it. Finally, in the third scenario (where a massive opening is proposed), we observe a \Ri{more pronounced increase} in case numbers, which will eventually exceed the number of infections observed at the beginning of the pandemic. \Ri{Noteworthy, this is more or less the current (March 2021) national trend (see, \eg \cite{Worldometerscountries})}. In this scenario, it would be uncertain to predict occupancy levels of ICU beds in hospitals as well as requests for mechanical respirators. \Ri{Nonetheless}, we would expect a saturation of the public health system.

\section{Conclusions}

COVID-19 spreading \Ri{dynamics depends} on multiple factors, including the biological/epidemiological aspects of SARS-CoV-2, human behavior, governmental \Ri{interventions}, and \Ri{heterogeneities among} the affected population. Several research groups have statistically analyzed these factors during pandemic progression, and we carefully included them in our modelling. The multi-group SEIRD model used considers the heterogeneous distribution and dynamic displacement of the Chilean population, grouped in 16 regions, and also a \Ri{timeline of the different NPIs enacted by the government.}

Following the presented results, our model shows to be efficient to the adjustment of the raw data, overcoming challenges as discontinuities and high variability in it, generating simulations that are easy to interpret and project \Rii{with narrow confidence intervals}. The measures imposed for a particular region can affect other regions, to a greater or lesser extent depending on their interaction, thus highlighting the need for coordinated governmental actions to control the spread of COVID-19. We have shown that the multi-group SEIRD model presented in this work is a useful tool to represent the contribution of each region in a heterogeneously populated country and is helpful to forecast in the short term the evolution of the different population groups.

\section*{Conflict of Interest Statement}

The authors declare that the research was conducted in the absence of any commercial or financial relationships that could be construed as a potential conflict of interest.

\section*{Author Contributions}
DF: Conceptualization; Methodology; Software development; Formal Analysis; Investigation; Writing - Original draft preparation; Writing - Review \& Editing.
NL-K: Conceptualization; Formal Analysis; Investigation; Data Curation; Writing - Original draft preparation; Visualization; Writing - Review \& Editing.
AS-D: Conceptualization; Writing - Original draft preparation; Writing - Review \& Editing \& Project administration.
AO-N: Project administration; Funding resources; Writing - Review \& Editing \& Project administration.

\section*{Acknowledgements}
The authors gratefully acknowledge support from the Centre for Biotechnology and Bioengineering - CeBiB (PIA project FB0001, Conicyt, Chile). AS-D thanks PAI Programme (I7818010006). The authors sincerely thank Sebastian Contreras for his support and advice during this research. \Rii{Source code (MATLAB) to reproduce the figures of this paper (main text and Supplementary Materials) is provided as Supplementary File, together with raw parameter data (for $\beta_i$ and $\theta_i$, in xlsx format) used to calculate the statistical ranges reported in the tables.}

\section*{References}
%

\appendix
\section{List of symbols}

$X$ \hspace{1cm} Arbitrary variable for representing a generic fraction 

$n_i$ \hspace{1cm} Base number of members class $i$

$n_i^{T}$ \hspace{0.9cm} Effective number of members class $i$

$\alpha$ \hspace{1.1cm} Asymptomatic ratio of the population

$\xi$ \hspace{1.2cm} Extra factor of behavioral virulence of asymptomatic patients

$\Phi^{ij}$ \hspace{0.9cm} Fraction of class $i$ in class $j$

$p_i$ \hspace{1.1cm} Immunity ratio of newborns of class $i$

$\Lambda_i$ \hspace{1cm} Net population growth rate $i$

$d_i$ \hspace{1.1cm} Per capita base death rate of class $i$

$\beta_i$ \hspace{1.1cm} Spreading rate of the virus in class $i$

$\epsilon_i$ \hspace{1.1cm} Inverse of the incubation time in class $i$

$\gamma_i$ \hspace{1.1cm} Recovery rate of class $i$

$\theta_i$ \hspace{1.1cm} Pathogen induced death rate in class $i$

$\Phi$ \hspace{1.1cm} Interaction matrix

$T_k$ \hspace{1cm} $k$'th epidemiological period

$t_{i}^{k},\,t_f^{k}$ \hspace{0.6cm} First and last day of the $k$'th epidemiological period

$e_i$ \hspace{1.1cm} Fraction of people from region $i$ travelling to other region

$C_{ij}$ \hspace{0.9cm} Fraction of those individuals leaving region $i$ that travel to region $j$

$w_X^{i,k}$ \hspace{0.8cm} Weighting empirical factor for the $X_i$ variable in the $k$'th epidemiological period

\end{document}